\documentclass[%
 aip,
cp,  
 amsmath,amssymb,
reprint,
]{revtex4-2}
\usepackage{graphicx}
\usepackage{dcolumn}
\usepackage{bm}

\usepackage[utf8]{inputenc}
\usepackage[T1]{fontenc}
\usepackage{mathptmx} 

\begin{document}

\title{Design and Construction of an Affordable Adjustable Repetition-Rate Optical Frequency Comb}

\author{Christopher E. Latchford} 
 \email{christopher.latchford@gmail.com}
\author{Daniel L. Maser}%
 \email{Corresponding author: dmaser@conncoll.edu}
 
\affiliation{
  Department of Physics, Astronomy, and Geophysics, Connecticut College, \\ 270 Mohegan Ave., New London, Connecticut 06320, USA
}

\date{\today} 

\begin{abstract}
We designed and built an adjustable repetition-rate optical frequency comb using off-the-shelf components with several improvements to an established erbium fiber design. This design, built fully by an undergraduate, was assembled at a fraction of the cost of a commercial comb. This comb has a repetition rate of roughly 100 MHz with a tuning range of 3 MHz. A stabilization control loop for $f_{rep}$ achieved long-term drifts of < 1 Hz over several days. This comb will eventually be implemented for use in dual-comb spectroscopy.
\end{abstract}

\maketitle

\section{\label{sec:level1}Background and Motivation}

An optical frequency comb is a specialized laser, for which the Nobel Prize was awarded in 2005, with many applications throughout physics, from its use in oil fields to find methane leaks \cite{coburn2018regional} to its use as a breathalyzer to detect trace molecules indicating disease \cite{thorpe2008cavity}. This technology combines the precision of a continuous-wave laser with the spectral bandwidth of a white-light source \cite{RevModPhys.78.1279}. Optical frequency combs are distinct from the more traditional tunable continuous-wave (CW) laser because, rather than containing a single instantaneous frequency, an optical frequency comb spans a large bandwidth of frequencies \cite{droste2016optical}. These lasers also have the precision to be used to measure fundamental constants and in tests of fundamental physics, from new measurements of hydrogen \cite{RevModPhys.78.1297} to the refractive index of air\cite{Minoshima:00}. An optical frequency comb is composed of many different teeth, where each tooth can be thought of as a CW laser described by the equation $f_n=nf_{rep}+f_0$ \cite{ideguchi2017dual}, where $f_{rep}$ is the repetition rate of the comb, dictated by the length of the oscillator, and $f_0$ is the offset frequency of the comb. 

We plan to introduce our newly designed and constructed comb in conjunction with a previously built optical frequency comb for use in dual-comb spectroscopy, which requires both combs to be stabilized in $f_{rep}$ and $f_0$ \cite{newbury2010sensitivity}. Because $f_{rep}$ is an RF reference, its stability must be at the millihertz level. $f_0$ will be indirectly stabilized by optically beating comb teeth from the two different combs to a shared narrow-linewidth laser and phase-locking those beat frequencies to RF references. Such a locking scheme will establish mutual coherence between the combs in the 1560-nm region \cite{Coddington:16}. Our goal is to show that this technology, with demonstrated mutual coherence, is not only accessible to state-of-the-art facilities, but also to smaller-scale, liberal arts institutions. 

\section{Design and Construction}

\begin{figure}
\includegraphics[scale=0.32]{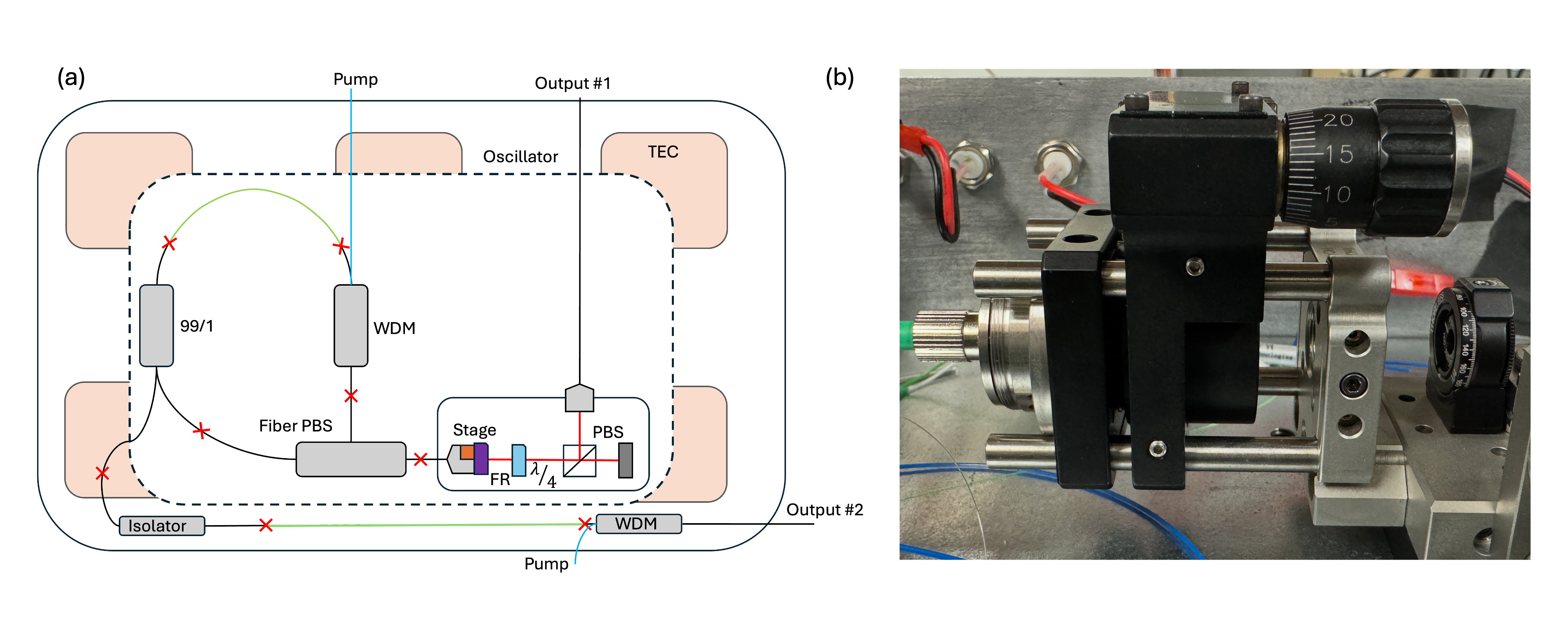}
\caption{\centering\label{fig:design} (a) A diagram of the "figure-9" design inside a sealed aluminum box. The $f_{rep}$ of the comb is roughly 100.037 MHz, which is achieved using roughly 2 m of fiber, 58 cm of which is EDF, and  14 cm of free space. 980-nm pump diodes drive both sections of EDF. Output 1 is the output of the free-space PBS, and output 2 is the 99/1 coupler output, which in the diagram is amplified in the EDFA. Red Xs indicate splices in fiber. (b) The adjustable free-space section, where the collimator position (left) can be adjusted coarsely by sliding the translation stage along the cage rails and finely using the translation stage's micrometer. The Faraday rotator (FR) is attached to the translation stage.}
\end{figure}

We based our design on the "figure-9" laser in Ref. [\onlinecite{hansel2018all}]. Our improvements to this design include the addition of a 99/1 coupler, a fiber polarizing beam splitter (PBS), and a z-axis translation stage, as shown in Fig. \ref{fig:design}(a). The second output provided by the 99/1 coupler allowed better metrics for alignment. The use of a fiber PBS, rather than free space, helped with free-space alignment. 

The z-axis translation stage, as seen in Fig. \ref{fig:design}(b), assists in achieving $f_{rep}$ near that of our existing, fixed $f_{rep}$ comb. In an all-fiber design, $f_{rep}$ precision is limited to the accuracy of the fiber cleaver and fiber length measurements, which is insufficient to reach $\Delta f_{rep}$ of 10 kHz (equivalent to roughly 0.2 mm of fiber), our target for dual-comb spectroscopy. In order to achieve higher precision, we introduced a free-space section, with several methods of free-space length adjustment. For both coarse and fine adjustment, we installed a z-axis translation stage along cage rails, where the stage can be slid along the rails for coarse adjustment of $f_{rep}$ over a length of 3 cm, and the stage’s micrometer can adjust $f_{rep}$ over a distance of 2 mm to better match our previously constructed comb. Our free-space-section, as seen in the inner box of Fig. \ref{fig:design}(a), consists of the z-axis translational stage with a micrometer attached, a Faraday rotator, a quarter-wave plate ($\lambda/4$), a PBS, and a mirror.
 
Outside of the oscillator we added an erbium-doped fiber amplifier (EDFA) into our box, consisting of erbium-doped fiber (EDF), a wavelength division multiplexer (WDM), and an isolator. This EDFA, seen at the bottom of Fig. \ref{fig:design}(a), is reverse-pumped where the pump and signal counterpropagate, allowing us to run lower power in our oscillator while maintaining strong output power. Including an EDFA also allows us to use either output as our main output to science, therefore using the other output as a tap output. 

In order to maintain mode-lock for extended periods of time as well as use in dual-comb spectroscopy, each comb needs $f_{rep}$ and $f_0$ stabilized. Many factors, such as temperature, vibrations, and electrical fluctuations, can cause both $f_{rep}$ and $f_0$ to drift. The dominant factor is temperature. As temperature drifts, the optical fibers will expand and contract, causing $f_{rep}$ to drift. We measured the drift of $f_{rep}$ using a high-speed photodetector. Our first step to counter these drifts is to enclose the comb in a sealed aluminum box, visible in Fig. \ref{fig:design}(b), which eliminates air currents and reduces temperature fluctuations. To actively combat these $f_{rep}$ changes, we have placed thermoelectric coolers (TEC) in our box underneath the oscillator. We use the TECs in a Python-based feedback loop tied to $f_{rep}$ directly rather than locking it to a specific temperature. As dual-comb spectroscopy requires a much tighter, high-bandwidth phase-lock of $f_{rep}$, we also installed two piezoelectric transducers (PZTs) to the control loop. We included one large PZT with a tunability of 20 $\mu$m, corresponding to changes of up to 100 Hz, and a small PZT with a tunability of 2 $\mu$m, corresponding to changes of 10 Hz. The PZTs, in tandem with the TECs, can maintain phase-lock amid environmental changes, as the TECs can correct for any large-scale long-term drift. We epoxied the fiber to the two PZTs and stabilized the fiber with modeling clay to negate high-frequency vibrations from the PZTs, based off the use of PZTs in Ref. [\onlinecite{sinclair2015invited}]. A summary of the $f_{rep}$ tuning components is given in Table 1.

\begin{center}
\begin{table}[htbp]
\caption{\label{tab:table1}Tunability, speed, and accuracy of components used in $f_{rep}$ control loop.}
\begin{tabular}{|c|c|c|c|} 
 \hline
  \textbf{Component} & \textbf{Tunability Range} & \textbf{Response Time} & \textbf{Accuracy} \\ [0.5ex] 
 \hline
 z-axis translational stage & 3 MHz & Manual & kHz \\
 \hline
 Micrometer & 100 kHz & Manual & 100 Hz \\
 \hline
 TECs & 100 kHz & Minutes & Hz \\ 
 \hline
 Large PZT & 100 Hz & 10 ms & mHz \\
 \hline
 Small PZT & 10 Hz & <1 ms & mHz \\[1ex] 
 \hline
\end{tabular}
\end{table}
\end{center}

\section{Results}

To initially test our design, we added extra patch cords to extend the length of the oscillator, first by 2 m to achieve $f_{rep}$ of 60 MHz, then by 1 m to achieve $f_{rep}$ of 80 MHz. The extra patch cords boosted the pulse energy, making the comb easier to mode-lock. We added these patch cords not only to test the design but to find the angles on the quarter-wave plate where the comb would mode-lock at the lowest pump power. After finding the angle (between 60\textdegree and 64\textdegree), we cut and spliced the comb to its final length.

Two 980-nm pump diodes, combined with a fiber PBS, provided a power of 726 mW. At higher power, our comb mode-locked with multipulsing, where, rather than forming a single pulse, the oscillator supports multiple evenly-spaced pulses per roundtrip, which appears as a modulated spectrum on an optical spectrum analyzer. Once the comb mode-locked with multipulsing, we lowered the power until it single-pulsed with CW breakthrough. To eliminate CW breakthrough, we further reduced the power until it was mode-locked without breakthrough, at roughly 300 mW of pump power, and remained mode-locked down to roughly 200 mW of pump power.

\begin{figure}[tbp]
\includegraphics[scale=0.57]{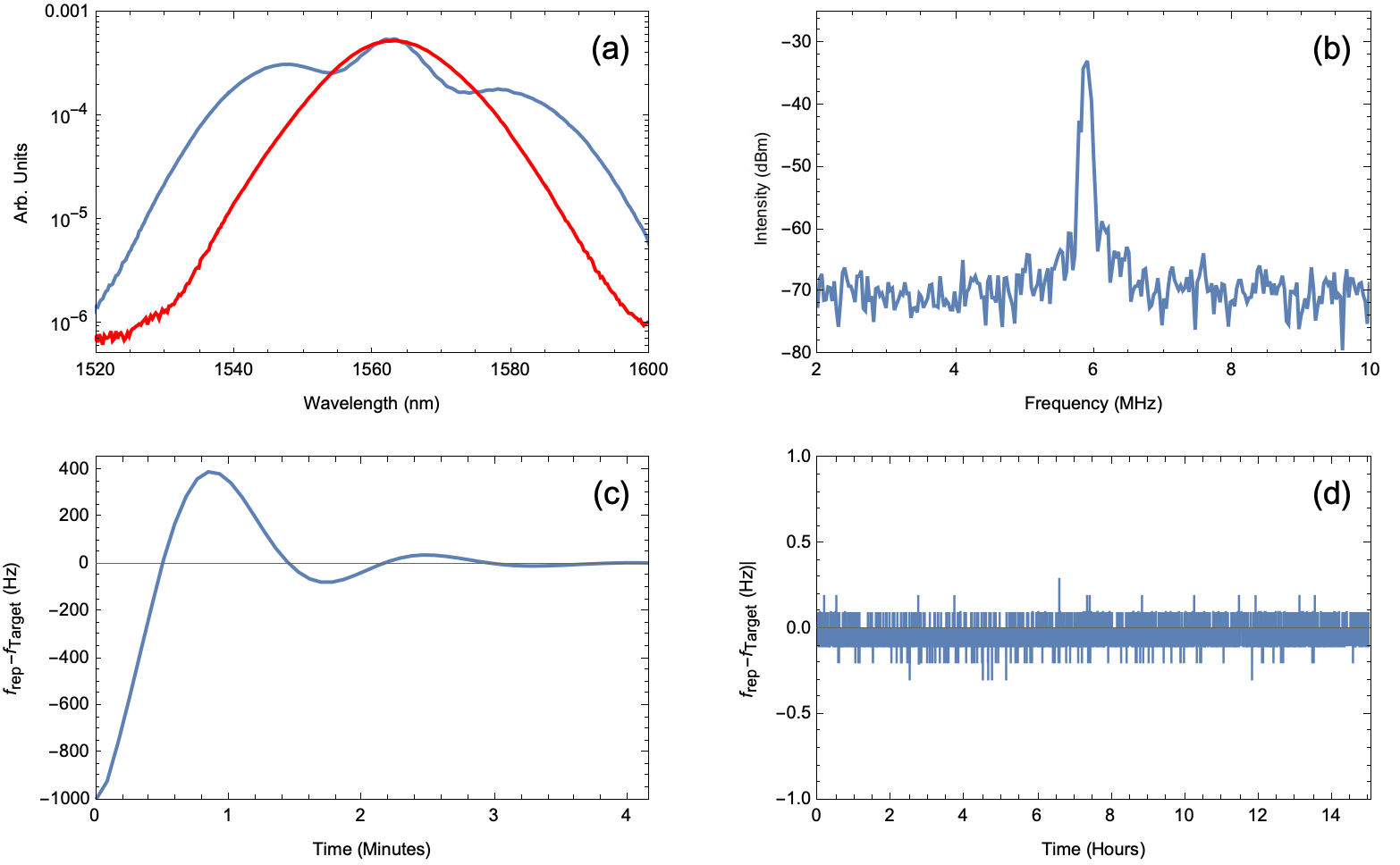}
\caption{\centering\label{fig:OSA_Spectrum} (a) Spectral output from the comb recorded on an optical spectrum analyzer with free-space output in blue and 99/1 coupler output in red. (b) Heterodyne beat between comb tooth and narrow-linewidth laser. The measurement was taken with a resolution bandwidth of 30 kHz. (c) $f_{rep}$ control loop performance, demonstrating the system's ability to reach a target $f_{rep}$ within minutes. (d) Long-term stabilization of $f_{rep}$, showing drifts of only 0.2 Hz over the course of 15 h. (c) and (d) were both measured using a HP 5385A frequency counter with 0.1-Hz resolution.}
\end{figure}

Figure \ref{fig:OSA_Spectrum}(a) displays spectra of both the 99/1 coupler output and the free-space PBS output, measured using an optical spectrum analyzer. Spectral broadening in the output fiber is observed in the unamplified free-space PBS output. The resulting spectrum is less suitable for our test spectroscopy near 1560 nm but is appropriate for later use with highly nonlinear fiber (HNLF). Based on the spectra of both the 99/1 output and the free-space PBS output, the free-space PBS output is currently better suited as a monitor and the 99/1 is more suitable for spectroscopy at 1560 nm. Based on its suitability, the power numbers reported refer to the 99/1 coupler output. At a pump power of roughly 200 mW, the output is approximately 5 $\mu$W, but with the EDFA pumped with 150 mW of 980-nm light, we achieved an output power of 10 mW. At this power, the EDFA did not broaden the spectrum from what is shown in Fig. \ref{fig:OSA_Spectrum}(a).

Once mode-locked, we were able to use the micrometer to achieve our goal of a 10-kHz difference between this comb and our preexisting comb. The comb is able to reach a target $f_{rep}$ within minutes, as shown in Fig. \ref{fig:OSA_Spectrum}(c), and remain at that target with a drift of under $\pm0.2$ Hz, as shown in Fig. \ref{fig:OSA_Spectrum}(d). This stability is achieved using only the TECs inside the box. To achieve $f_{rep}$ stabilization in the millihertz realm, our next step is to include the PZTs in our control loop to phase-lock the comb. We also recorded the heterodyne beat between a single comb tooth and a narrow-linewidth laser, as displayed in Fig. \ref{fig:OSA_Spectrum}(b), thus showing the structure of the oscillator. We estimate a 20-dB linewidth of 250 kHz, as expected for a free-running erbium fiber comb \cite{droste2016optical}. The heterodyne beat also demonstrates the suitability to phase-lock the beat to an RF reference, which will indirectly stabilize $f_0$ as described in the Background and Motivation section. 

\section{Conclusion}

We were able to design and construct an adjustable $f_{rep}$ optical frequency comb for a fraction of the cost of a commercial system at roughly \$5,000 in parts. We were able to build our comb using off-the-shelf components while also having an undergraduate splice, align, and mode-lock the entirety of this project. Our next goal is to use this comb in conjunction with our preexisting comb to collect interferograms of the dual-comb spectra of hydrogen cyanide (HCN), which has absorption features centered in the spectral coverage of both our combs. HCN will be a good test of the suitability of our combs for dual-comb spectroscopy. 

\begin{acknowledgments}
This work was supported by donors of the ACS Petroleum Research Fund under Undergraduate New Investigator Grant No. 65007-UNI6. We acknowledge assistance in crimping and soldering electrical connections from Michael Moran and machining the enclosure from Tom Dobkowski, and we also thank Professor Jacob Stewart for useful feedback on this manuscript. 
\end{acknowledgments}

\nocite{*}
\bibliography{aipsamp}

\providecommand{\noopsort}[1]{}\providecommand{\singleletter}[1]{#1}%
\begin{thebibliography}{11}%
\makeatletter
\providecommand \@ifxundefined [1]{%
 \@ifx{#1\undefined}
}%
\providecommand \@ifnum [1]{%
 \ifnum #1\expandafter \@firstoftwo
 \else \expandafter \@secondoftwo
 \fi
}%
\providecommand \@ifx [1]{%
 \ifx #1\expandafter \@firstoftwo
 \else \expandafter \@secondoftwo
 \fi
}%
\providecommand \natexlab [1]{#1}%
\providecommand \enquote  [1]{``#1''}%
\providecommand \bibnamefont  [1]{#1}%
\providecommand \bibfnamefont [1]{#1}%
\providecommand \citenamefont [1]{#1}%
\providecommand \href@noop [0]{\@secondoftwo}%
\providecommand \href [0]{\begingroup \@sanitize@url \@href}%
\providecommand \@href[1]{\@@startlink{#1}\@@href}%
\providecommand \@@href[1]{\endgroup#1\@@endlink}%
\providecommand \@sanitize@url [0]{\catcode `\\12\catcode `\$12\catcode `\&12\catcode `\#12\catcode `\^12\catcode `\_12\catcode `\%12\relax}%
\providecommand \@@startlink[1]{}%
\providecommand \@@endlink[0]{}%
\providecommand \url  [0]{\begingroup\@sanitize@url \@url }%
\providecommand \@url [1]{\endgroup\@href {#1}{\urlprefix }}%
\providecommand \urlprefix  [0]{URL }%
\providecommand \Eprint [0]{\href }%
\providecommand \doibase [0]{http://dx.doi.org/}%
\providecommand \selectlanguage [0]{\@gobble}%
\providecommand \bibinfo  [0]{\@secondoftwo}%
\providecommand \bibfield  [0]{\@secondoftwo}%
\providecommand \translation [1]{[#1]}%
\providecommand \BibitemOpen [0]{}%
\providecommand \bibitemStop [0]{}%
\providecommand \bibitemNoStop [0]{.\EOS\space}%
\providecommand \EOS [0]{\spacefactor3000\relax}%
\providecommand \BibitemShut  [1]{\csname bibitem#1\endcsname}%
\let\auto@bib@innerbib\@empty
\bibitem [{\citenamefont {Coburn}\ \emph {et~al.}(2018)\citenamefont {Coburn}, \citenamefont {Alden}, \citenamefont {Wright}, \citenamefont {Cossel}, \citenamefont {Baumann}, \citenamefont {Truong}, \citenamefont {Giorgetta}, \citenamefont {Sweeney}, \citenamefont {Newbury}, \citenamefont {Prasad} \emph {et~al.}}]{coburn2018regional}%
  \BibitemOpen
  \bibfield  {author} {\bibinfo {author} {\bibfnamefont {S.}~\bibnamefont {Coburn}}, \bibinfo {author} {\bibfnamefont {C.~B.}\ \bibnamefont {Alden}}, \bibinfo {author} {\bibfnamefont {R.}~\bibnamefont {Wright}}, \bibinfo {author} {\bibfnamefont {K.}~\bibnamefont {Cossel}}, \bibinfo {author} {\bibfnamefont {E.}~\bibnamefont {Baumann}}, \bibinfo {author} {\bibfnamefont {G.-W.}\ \bibnamefont {Truong}}, \bibinfo {author} {\bibfnamefont {F.}~\bibnamefont {Giorgetta}}, \bibinfo {author} {\bibfnamefont {C.}~\bibnamefont {Sweeney}}, \bibinfo {author} {\bibfnamefont {N.~R.}\ \bibnamefont {Newbury}}, \bibinfo {author} {\bibfnamefont {K.}~\bibnamefont {Prasad}},  \emph {et~al.},\ }\bibfield  {title} {\enquote {\bibinfo {title} {Regional trace-gas source attribution using a field-deployed dual frequency comb spectrometer},}\ }\href@noop {} {\bibfield  {journal} {\bibinfo  {journal} {Optica}\ }\textbf {\bibinfo {volume} {5}},\ \bibinfo {pages} {320--327} (\bibinfo {year} {2018})}\BibitemShut {NoStop}%
\bibitem [{\citenamefont {Thorpe}\ \emph {et~al.}(2008)\citenamefont {Thorpe}, \citenamefont {Balslev-Clausen}, \citenamefont {Kirchner},\ and\ \citenamefont {Ye}}]{thorpe2008cavity}%
  \BibitemOpen
  \bibfield  {author} {\bibinfo {author} {\bibfnamefont {M.~J.}\ \bibnamefont {Thorpe}}, \bibinfo {author} {\bibfnamefont {D.}~\bibnamefont {Balslev-Clausen}}, \bibinfo {author} {\bibfnamefont {M.~S.}\ \bibnamefont {Kirchner}}, \ and\ \bibinfo {author} {\bibfnamefont {J.}~\bibnamefont {Ye}},\ }\bibfield  {title} {\enquote {\bibinfo {title} {Cavity-enhanced optical frequency comb spectroscopy: Application to human breath analysis},}\ }\href@noop {} {\bibfield  {journal} {\bibinfo  {journal} {Opt. Express}\ }\textbf {\bibinfo {volume} {16}},\ \bibinfo {pages} {2387--2397} (\bibinfo {year} {2008})}\BibitemShut {NoStop}%
\bibitem [{\citenamefont {Hall}(2006)}]{RevModPhys.78.1279}%
  \BibitemOpen
  \bibfield  {author} {\bibinfo {author} {\bibfnamefont {J.~L.}\ \bibnamefont {Hall}},\ }\bibfield  {title} {\enquote {\bibinfo {title} {Nobel lecture: Defining and measuring optical frequencies},}\ }\href {\doibase 10.1103/RevModPhys.78.1279} {\bibfield  {journal} {\bibinfo  {journal} {Rev. Mod. Phys.}\ }\textbf {\bibinfo {volume} {78}},\ \bibinfo {pages} {1279--1295} (\bibinfo {year} {2006})}\BibitemShut {NoStop}%
\bibitem [{\citenamefont {Droste}\ \emph {et~al.}(2016)\citenamefont {Droste}, \citenamefont {Ycas}, \citenamefont {Washburn}, \citenamefont {Coddington},\ and\ \citenamefont {Newbury}}]{droste2016optical}%
  \BibitemOpen
  \bibfield  {author} {\bibinfo {author} {\bibfnamefont {S.}~\bibnamefont {Droste}}, \bibinfo {author} {\bibfnamefont {G.}~\bibnamefont {Ycas}}, \bibinfo {author} {\bibfnamefont {B.~R.}\ \bibnamefont {Washburn}}, \bibinfo {author} {\bibfnamefont {I.}~\bibnamefont {Coddington}}, \ and\ \bibinfo {author} {\bibfnamefont {N.~R.}\ \bibnamefont {Newbury}},\ }\bibfield  {title} {\enquote {\bibinfo {title} {Optical frequency comb generation based on erbium fiber lasers},}\ }\href@noop {} {\bibfield  {journal} {\bibinfo  {journal} {Nanophotonics}\ }\textbf {\bibinfo {volume} {5}},\ \bibinfo {pages} {196--213} (\bibinfo {year} {2016})}\BibitemShut {NoStop}%
\bibitem [{\citenamefont {H\"ansch}(2006)}]{RevModPhys.78.1297}%
  \BibitemOpen
  \bibfield  {author} {\bibinfo {author} {\bibfnamefont {T.~W.}\ \bibnamefont {H\"ansch}},\ }\bibfield  {title} {\enquote {\bibinfo {title} {Nobel lecture: Passion for precision},}\ }\href {\doibase 10.1103/RevModPhys.78.1297} {\bibfield  {journal} {\bibinfo  {journal} {Rev. Mod. Phys.}\ }\textbf {\bibinfo {volume} {78}},\ \bibinfo {pages} {1297--1309} (\bibinfo {year} {2006})}\BibitemShut {NoStop}%
\bibitem [{\citenamefont {Minoshima}\ and\ \citenamefont {Matsumoto}(2000)}]{Minoshima:00}%
  \BibitemOpen
  \bibfield  {author} {\bibinfo {author} {\bibfnamefont {K.}~\bibnamefont {Minoshima}}\ and\ \bibinfo {author} {\bibfnamefont {H.}~\bibnamefont {Matsumoto}},\ }\bibfield  {title} {\enquote {\bibinfo {title} {High-accuracy measurement of 240-m distance in an optical tunnel by use of a compact femtosecond laser},}\ }\href {\doibase 10.1364/AO.39.005512} {\bibfield  {journal} {\bibinfo  {journal} {Appl. Opt.}\ }\textbf {\bibinfo {volume} {39}},\ \bibinfo {pages} {5512--5517} (\bibinfo {year} {2000})}\BibitemShut {NoStop}%
\bibitem [{\citenamefont {Ideguchi}(2017)}]{ideguchi2017dual}%
  \BibitemOpen
  \bibfield  {author} {\bibinfo {author} {\bibfnamefont {T.}~\bibnamefont {Ideguchi}},\ }\bibfield  {title} {\enquote {\bibinfo {title} {Dual-comb spectroscopy},}\ }\href@noop {} {\bibfield  {journal} {\bibinfo  {journal} {Opt. Photonics News}\ }\textbf {\bibinfo {volume} {28}},\ \bibinfo {pages} {32--39} (\bibinfo {year} {2017})}\BibitemShut {NoStop}%
\bibitem [{\citenamefont {Newbury}, \citenamefont {Coddington},\ and\ \citenamefont {Swann}(2010)}]{newbury2010sensitivity}%
  \BibitemOpen
  \bibfield  {author} {\bibinfo {author} {\bibfnamefont {N.~R.}\ \bibnamefont {Newbury}}, \bibinfo {author} {\bibfnamefont {I.}~\bibnamefont {Coddington}}, \ and\ \bibinfo {author} {\bibfnamefont {W.}~\bibnamefont {Swann}},\ }\bibfield  {title} {\enquote {\bibinfo {title} {Sensitivity of coherent dual-comb spectroscopy},}\ }\href@noop {} {\bibfield  {journal} {\bibinfo  {journal} {Opt. Express}\ }\textbf {\bibinfo {volume} {18}},\ \bibinfo {pages} {7929--7945} (\bibinfo {year} {2010})}\BibitemShut {NoStop}%
\bibitem [{\citenamefont {Coddington}, \citenamefont {Newbury},\ and\ \citenamefont {Swann}(2016)}]{Coddington:16}%
  \BibitemOpen
  \bibfield  {author} {\bibinfo {author} {\bibfnamefont {I.}~\bibnamefont {Coddington}}, \bibinfo {author} {\bibfnamefont {N.}~\bibnamefont {Newbury}}, \ and\ \bibinfo {author} {\bibfnamefont {W.}~\bibnamefont {Swann}},\ }\bibfield  {title} {\enquote {\bibinfo {title} {Dual-comb spectroscopy},}\ }\href {\doibase 10.1364/OPTICA.3.000414} {\bibfield  {journal} {\bibinfo  {journal} {Optica}\ }\textbf {\bibinfo {volume} {3}},\ \bibinfo {pages} {414--426} (\bibinfo {year} {2016})}\BibitemShut {NoStop}%
\bibitem [{\citenamefont {H{\"a}nsel}\ \emph {et~al.}(2017)\citenamefont {H{\"a}nsel}, \citenamefont {Hoogland}, \citenamefont {Giunta}, \citenamefont {Schmid}, \citenamefont {Steinmetz}, \citenamefont {Doubek}, \citenamefont {Mayer}, \citenamefont {Dobner}, \citenamefont {Cleff}, \citenamefont {Fischer},\ and\ \citenamefont {Holzwarth}}]{hansel2018all}%
  \BibitemOpen
  \bibfield  {author} {\bibinfo {author} {\bibfnamefont {W.}~\bibnamefont {H{\"a}nsel}}, \bibinfo {author} {\bibfnamefont {H.}~\bibnamefont {Hoogland}}, \bibinfo {author} {\bibfnamefont {M.}~\bibnamefont {Giunta}}, \bibinfo {author} {\bibfnamefont {S.}~\bibnamefont {Schmid}}, \bibinfo {author} {\bibfnamefont {T.}~\bibnamefont {Steinmetz}}, \bibinfo {author} {\bibfnamefont {R.}~\bibnamefont {Doubek}}, \bibinfo {author} {\bibfnamefont {P.}~\bibnamefont {Mayer}}, \bibinfo {author} {\bibfnamefont {S.}~\bibnamefont {Dobner}}, \bibinfo {author} {\bibfnamefont {C.}~\bibnamefont {Cleff}}, \bibinfo {author} {\bibfnamefont {M.}~\bibnamefont {Fischer}}, \ and\ \bibinfo {author} {\bibfnamefont {R.}~\bibnamefont {Holzwarth}},\ }\bibfield  {title} {\enquote {\bibinfo {title} {All polarization-maintaining fiber laser architecture for robust femtosecond pulse generation},}\ }\href {\doibase 10.1007/s00340-016-6598-2} {\bibfield  {journal} {\bibinfo  {journal} {Appl. Phys. B}\ }\textbf {\bibinfo {volume} {123}},\ \bibinfo
  {pages} {41} (\bibinfo {year} {2017})}\BibitemShut {NoStop}%
\bibitem [{\citenamefont {Sinclair}\ \emph {et~al.}(2015)\citenamefont {Sinclair}, \citenamefont {Desch{\^e}nes}, \citenamefont {Sonderhouse}, \citenamefont {Swann}, \citenamefont {Khader}, \citenamefont {Baumann}, \citenamefont {Newbury},\ and\ \citenamefont {Coddington}}]{sinclair2015invited}%
  \BibitemOpen
  \bibfield  {author} {\bibinfo {author} {\bibfnamefont {L.~C.}\ \bibnamefont {Sinclair}}, \bibinfo {author} {\bibfnamefont {J.-D.}\ \bibnamefont {Desch{\^e}nes}}, \bibinfo {author} {\bibfnamefont {L.}~\bibnamefont {Sonderhouse}}, \bibinfo {author} {\bibfnamefont {W.~C.}\ \bibnamefont {Swann}}, \bibinfo {author} {\bibfnamefont {I.~H.}\ \bibnamefont {Khader}}, \bibinfo {author} {\bibfnamefont {E.}~\bibnamefont {Baumann}}, \bibinfo {author} {\bibfnamefont {N.~R.}\ \bibnamefont {Newbury}}, \ and\ \bibinfo {author} {\bibfnamefont {I.}~\bibnamefont {Coddington}},\ }\bibfield  {title} {\enquote {\bibinfo {title} {Invited article: A compact optically coherent fiber frequency comb},}\ }\href@noop {} {\bibfield  {journal} {\bibinfo  {journal} {Rev. Sci. Instrum.}\ }\textbf {\bibinfo {volume} {86}},\ \bibinfo {pages} {081301} (\bibinfo {year} {2015})}\BibitemShut {NoStop}%
\end{thebibliography}%

\end{document}